\documentclass[twocolumn,prl,showpacs,preprintnumbers]{revtex4}

\usepackage{amsmath} 
\usepackage{amsfonts} 
\usepackage{amssymb}
\usepackage{graphicx} 
\usepackage{psfrag}

\newcommand{\half}{\tfrac{1}{2}}

\newcommand{\g}{{g}}
\newcommand{\Mp}{{M_{\rm P}}}

\newcommand{\Gn}{{G_{\rm Newton}}}

\begin{document}
\preprint{MIT-CTP-3617}
\preprint{hep-th/0509050}

\title{Gravitational Correction to Running of Gauge Couplings}
\author{Sean P.~Robinson}
\email{spatrick@mit.edu}
\author{Frank Wilczek}
\email{wilczek@mit.edu}
\affiliation{ Center for Theoretical Physics,
    Laboratory for Nuclear Science and Department of Physics,
    Massachusetts Institute of Technology,  Cambridge, Massachusetts 02139 }

\begin{abstract} 
We calculate the contribution of graviton exchange to the running of gauge couplings at lowest non-trivial order in perturbation theory.  Including this contribution in a theory that features coupling constant unification does not upset this unification, but 
rather shifts the unification scale.   When extrapolated formally, the gravitational correction renders all gauge couplings asymptotically free.
\end{abstract}
\maketitle

The straightforward framework for quantum gravity---general relativity  quantized for small fluctuations around flat space---is a famously nonrenormalizable quantum field theory \cite{'tHooft:1974bx,Deser:1974cz,Deser:1974cy,Deser:1974xq}.   Nevertheless, this framework is appropriate for describing interactions at energies and momenta below the Planck scale $\Mp \equiv \sqrt{\hbar c /\Gn}\approx 1.4\times10^{19}$ GeV$/c^2$ when treated as an effective low-energy theory.   Indeed, if one makes subtractions to normalize physical couplings at an energy scale $E_0$ well below $\Mp$ in such a way as to enforce the Einstein-Hilbert action of general relativity at the classical level with minimal couplings and a vanishing (or very small) cosmological term, then quantum corrections to this classical action at scale $E$ will occur with coefficients containing positive powers of $(E-E_0)/\Mp$, a small number.    That procedure is the implicit foundation for practical use of classical general relativity as a model of nature despite the existence of quantum mechanics.  It therefore underlies an enormous range of successful physical and astrophysical applications. Only the classical theory really comes into play in those applications, because the quantum  corrections are quantitatively small.    Thus, the  conceptual framework  of effective field theory provides a sophisticated rationalization for  proceeding naively in applying the classical theory.

Still, as Donoghue has emphasized \cite{Donoghue:1994dn}, calculating  corrections to the classical theory is a problem of methodological interest.   Moreover, quantitative considerations concerning interactions at ultra-high energy scales, perhaps approaching the Planck scale, are important in assessing the possibility of gauge theory coupling unification \cite{Georgi:1973, Dimopoulos:1981yj}.   Also, the size of gravitational corrections, in comparison to the leading classical term, give an objective indication for the characteristic scale for the onset of quantum gravity phenomenology.  With these motivations, we consider here the one-loop (that is, first non-trivial order in perturbation theory) gravitational correction to running of gauge theory couplings.

We will perform this calculation directly in the framework described above. Any would-be fundamental theory of quantum gravity should reproduce the same result in the limit of the physical scenario considered here, which is bosonic gravity in a four dimensional Minkowski background, with general matter and gauge sectors, at energies below the Plank scale. Related calculations have been done in string theory \cite{Kiritsis:1994ta,Kiritsis:1997hj}, but this brings in several additional structures simultaneously, and we have found the results difficult to compare. 

\begin{figure}
\psfrag{YM}{}
\includegraphics[width=2in]{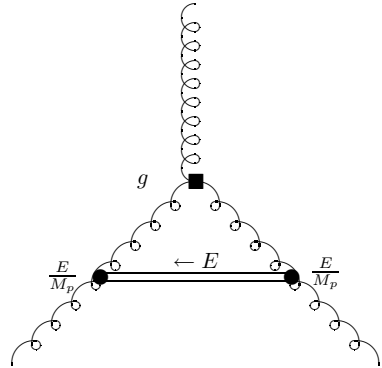}
\caption{\label{fig1} A typical Feynman diagram for a gravitational process contributing to the renormalization of a gauge coupling at one-loop. Curly lines represent gluons. Double lines represent gravitons. The three-gluon vertex $\blacksquare$ is proportional to $\g$, while the gluon-graviton vertex $\bullet$ is proportional to $E/\Mp$.}
\end{figure}

{\it Form of the correction}.---The character of the correction can be determined on very general grounds.  
The one-loop Feynman diagrams of interest involve a gluon 
vertex dressed by graviton exchange (See Fig. \ref{fig1}).
Alternatively, one could calculate the running coupling of a gluon to a test ``matter'' field.   Gauge invariance (i.e., universality of the gauge coupling) implies that the same result must be obtained.   This consideration highlights a cancellation between vertex and wave function renormalization, guaranteed by Ward identities, as is familiar in QED.   

Since the gauge boson vertex has
strength $\g$ and gravitons couple to energy-momentum with a dimensional coupling $\propto 1/\Mp$, dimensional analysis implies that the running of couplings in four dimensions will be governed by a Callan-Symanzik $\beta$ function 
of the form
\begin{equation}
\beta(\g,E)\equiv\frac{d\g}{d\ln{E}} = -\frac{b_0}{(4\pi)^2} \g^3 + a_0 \frac{E^2}{\Mp^2}\g,
\label{beta}
\end{equation}
where the first term includes the familiar non-gravitational contribution, and the second term includes the gravitational contribution. Since gravitons carry no gauge charge, $b_0$ has the same value it had in the absence of gravitation, as determined by the matter content and the gauge group.   
Detailed calculation is required to determine the numerical value of the coefficient $a_0$.

Even before knowing the value of $a_0$, much phenomenology can be extracted from the form of Eq. (\ref{beta}). The equation can be integrated to yield
\begin{multline}
e^{a_0 E^2/\Mp^2}\frac{(4\pi)^2}{\g(E)^2} - e^{a_0 E_0^2/\Mp^2}\frac{(4\pi)^2}{\g(E_0)^2} \\ =
b_0\int^{E^2}_{E_0^2} {e^{a_0 y / \Mp^2} \frac{dy}{y}}.
\label{formalSolution}
\end{multline}
This reduces, of course, to the familiar logarithmic running of inverse couplings in the limit $a_0 \rightarrow 0$ (or $\Mp \rightarrow \infty$).  Corrections to the familiar result will be very small for $E, E_0  \ll \Mp$.  

The value of $a_0$ is manifestly independent of the gauge interaction involved.   So, if we consider several gauge couplings $\g^i$, with different $b_0^i$, the condition that they unify at a common value $\g(E_{\rm U})$ is 
\begin{equation}
\frac{\frac{(4\pi)^2}{\g^i(E_0)^2}  - \frac{(4\pi)^2}{\g^j(E_0)^2}}{b_0^j - b_0^i} = 
e^{-a_0 E_0^2 / \Mp^2}\int^{E_{\rm U}^2}_{E_0^2} e^{a_0 y / \Mp^2} \frac{dy}{y}
\label{unify}
\end{equation}
for all choices of $i$ and $j$. In particular, the left-hand side of Eq. (\ref{unify}) must be independent of the choice of $i, j$.  Since that combination of initial couplings and renormalization group coefficients is not affected by the gravitational correction, the unification constraints remain unchanged.  However, the scale of unification and the value of the common coupling at unification do change, as we shall now discuss.   

Comparing the unification condition with and without the gravitational correction, and taking $E_0^2/\Mp^2 \rightarrow 0$, we see that the relationship between the uncorrected unification energy $E_*$ and the corrected unification energy $E_{\rm U}$ is
\begin{equation}
\ln{E_*^2 }= \ln{E_{\rm U}^2} + a_0 \frac{E_{\rm U}^2}{\Mp^2}.
\end{equation}
If $E_* \ll \Mp$, the self-consistent correction is
\begin{equation}
E_{\rm U}^2 \approx E_*^2\left (1 - a_0 \frac{E_*^2}{\Mp^2}\right). \label{ecorr}
\end{equation}
In standard (quasiminimal) unification schemes we find that $E_*$ is smaller than $\Mp$ by roughly 3 orders of magnitude, so this approximation is appropriate, but the correction itself is of no practical importance.   On the other hand, it is widely viewed as disturbing to have the separation of scales $E_* \ll \Mp$. Theories that address this issue will inevitably bring in the gravitational correction---which, if $a_0$ happens to be negative, helps to close the gap.  

The value of the coupling at unification is modified according to
\begin{equation}
e^{a_0 E_{\rm U}^2 / \Mp^2} \frac {1}{\g(E_{\rm U})^2} = \frac {1}{\g_*(E_*)^2},
\label{gmod}
\end{equation}
where $\g_*$  is the running coupling as determined by Eq. (\ref{beta}) with $a_0 \rightarrow 0$. For $E_* \ll \Mp$,
\begin{equation}
g(E_{\rm U})^2 \approx\left (1+ a_0 \frac{E_*^2}{\Mp^2}\right) g_*(E_*)^2.
\label{gcorr}
\end{equation}

{\it Method of calculation}.---The algebra required to evaluate $a_0$ in any straightforward way is formidable.   We document algebraic details elsewhere \cite{calculation}; here we just sketch our method and conventions.   

The dynamics for a non-Abelian gauge field coupled to gravity in 
$3+1$ spacetime dimensions is given by
the action
\begin{equation} 
S[{\bf g},{\cal A}]=\int{d^4x\sqrt{-{\bf g}}\left[\frac{1}{\kappa^2}{\bf R}
            -\frac{1}{4\g^2}  
            {\bf g}^{ac}{\bf g}^{bd}{\cal F}^{\bf a}_{ab}{\cal F}^{\bf a}_{cd}\right]},
\label{e2.0.1}
\end{equation} 
where ${\bf g}=\det{\bf g}_{ab}$, ${\bf g}_{ab}$ is the spacetime metric, $\kappa^2 = 16\pi/\Mp^2$, ${\bf R}$ is 
the Ricci scalar, $\g$ is the gauge coupling,
\begin{equation}
{\cal F}^{\bf a}_{ab} \equiv 
                 \Delta_a {\cal A}^{\bf a}_b-\Delta_b{\cal A}^{\bf a}_a
                  +f^{\bf abc}{\cal A}^{\bf b}_a{\cal A}^{\bf c}_b
\label{e2.0.2}
\end{equation}
is the field strength, ${\cal A}^{\bf a}_a$ is the gauge field,  $f^{\bf abc}$
are the structure constants of the non-Abelian gauge group $G$, and $\Delta_a$ is 
the spacetime covariant derivative operator.  Since ${\cal F}^{\bf a}_{ab}$ is antisymmetric under 
$a\leftrightarrow b$, the Christoffel connections arising from the derivatives
 in Equation (\ref{e2.0.2}) cancel against each other, so the covariant derivatives here can be 
replaced with ordinary derivatives. 

We apply the background field method because it is especially well suited to the specific problem at hand in ways that we will highlight throughout this section. In accordance with this method, we seek to evaluate the effective action for classical field configurations by integrating over quantum fluctuations $h_{ab}$ and $A^{\bf a}_a$: 
\begin{equation}
e^{iS_{\rm eff}[g,a]} =\int{{\cal D}h {\cal D} A e^{iS[{\bf g} ,{\cal A}] }}. \label{effact}
\end{equation}
Here, we have expanded ${\bf g}_{ab}$ as a quantum fluctuation $h_{ab}$
about a background $g_{ab}$,
\begin{equation}
{\bf g}_{ab} =g_{ab} + h_{ab} ,
\label{e2.0.3}
\end{equation}
and likewise expanded ${\cal A}^{\bf a}_a$  as a fluctuation  $A^{\bf a}_a$ about  a background  $a^{\bf a}_a$,
\begin{equation}
{\cal A}^{\bf a}_a  = a^{\bf a}_a + A^{\bf a}_a  \label{e2.0.4}.
\end{equation}
In principle, the classical fields $g_{ab}$ and $a^{\bf a}_a$ could satisfy the classical equations of motion, the coupled Einstein-Yang-Mills equations. For our purposes, however---that is, calculating the renormalization of gauge couplings to one-loop order in perturbation theory---it suffices to set $g_{ab}$ equal to the flat Minkowski metric while allowing $a^{\bf a}_a$ to obey the flat-space Yang-Mills equations of motion.  
We expand the action (\ref{e2.0.1}) in terms of these backgrounds and fluctuations up to quadratic order in 
the quantum fields, since within the background field method, higher-order terms in the action will only contribute to higher-loop 
processes. 

If we couple any matter to this system, we do not expand the matter fields as fluctuations about a background because we are only interested in the renormalization of the gauge coupling.  We still keep terms up to only quadratic order in quantum fields, however. Since we are expanding about Minkowski spacetime,  matter terms in the action will give exactly the same contribution to the one-loop renormalized coupling that they did in the absence of gravitation.  
 
The action (\ref{e2.0.1}) is invariant under diffeomorphisms
\begin{subequations} \label{e2.2.group1}
\begin{align}
\delta_\eta h_{ab}  =& \partial_a\eta_b + \partial_b \eta_a 
             +\partial_a\eta^ch_{cb}+\partial_b\eta^ch_{ca}+\eta^c\partial_c h_{ab}
\label{e2.2.1}, \\
\delta_\eta A^{\bf a}_a =&A^{\bf a}_c\partial_a\eta^c +\eta^c\partial_cA^{\bf a}_a
  \label{e2.2.2},  \\
\delta_\eta a^{\bf a}_a =&a^{\bf a}_c\partial_a\eta^c +\eta^c\partial_ca^{\bf a}_a
\label{e2.2.x};
\end{align}
\end{subequations}
and under gauge transformations of the group $G$
\begin{subequations} \label{e2.2.group2}
\begin{align}
\delta_\alpha A^{\bf a}_a  =& D_a\alpha^{\bf a}+f^{\bf abc}A^{\bf b}_a\alpha^{\bf c}
\label{e2.2.4},\\
\delta_\alpha h_{ab} =& 0
\label{e2.2.5}.
\end{align}
\end{subequations}
(Here, $D_a=\partial_a-ia^{\bf a}_a{\bf t}_r^{\bf a}$ when acting on representation $r$, and indices are raised and lowered with the background Minkowski metric.) These gauge symmetries need to fixed  before performing the functional integration (\ref{effact}). We 
take the background-covariant gauge-fixing conditions
\begin{align}
G^{\bf a}(A) \equiv& D_a A^{{\bf a}a} =0 \label{e2.2.12},\\
\tilde{C}^a(h,A) \equiv&C^a(h) -\frac{\kappa^2}{\g^2}F^{{\bf a}ab}A^{\bf a}_b=0 
\label{e2.2.11},
\end{align}
where 
\begin{equation}
C^a(h)\equiv\partial_b h^{ab}-\half \partial^a h \qquad(h\equiv h^a_a)
\label{e2.x.2},
\end{equation}
and $F^{\bf a}_{ab}$ is the appropriate function of classical fields only. Equation $(\ref{e2.2.11})$ is similar to an
$R_{\xi}$ gauge \cite{'tHooft:1972fi}.  Here it is engineered to cancel unpleasant 
graviton-gluon cross-terms that would otherwise appear in the expansion. 
Using the Faddeev-Popov method \cite{Popov:1967fd} in conjunction with Feynman-'t Hooft weighting 
factors, the gauge choices each add gauge-fixing terms to the action as well as corresponding ghost
fields. The ghost fields will not be expanded about a background and always appear at quadratic order in the action. So, like matter fields, the ghost fields contribute to the renormalized gauge couplings exactly as they do in the absence of gravitation. In particular, this means that the diffeomorphism ghost does not contribute at all and can be ignored in this background field calculation.

Not all gluon-graviton cross terms in the action can be eliminated by the choice of gauge (\ref{e2.2.11}) because gluon-graviton mixing in a vector background is a real physical effect. In order to evaluate the Gaussian integrals in Eq.~(\ref{effact}) as functional determinants, we formally combine $h_{ab}$ and $A^{\bf a}_a$ into a ``superfield'' such that the mixing terms appear in the off-diagonal entries of the functional matrix in question. If multiple gauge symmetries are present, each with its own gauge field and renormalizable coupling, the superfield must be expanded to include each gluon type, as well as the graviton. The functional matrix then contains cross terms that mix different gluon types, but these do not ultimately contribute to the calculation at one-loop order.  So, to this order, each gauge coupling gets renormalized independently.

{\it Result}.---At this point the Gaussian integrals over the quantum fields in Eq. (\ref{effact}) are formally defined, but the resulting functional determinants contain ultraviolet divergences.  We subtract them at a reference energy $E_0$.  We find the one-loop effective action at energy scale $E$ is
\begin{multline}
S_{\rm eff}[g,a]\approx
   -\frac{1}{4}\int{ d^4x
\left[\frac{1}{\g^2}+\frac{\kappa^2}{\g^2}\frac{3}{(4\pi)^2}(E^{2}-E_0^{2})
\right.}\\\left.+\frac{b_0}{(4\pi)^2}\ln{\frac{E^2}{E_0^2}}\right]F^{\bf a}_{ab}F^{{\bf a}ab}
\label{e3.19},
\end{multline}
where $b_0$ depends on the gauge and matter content independently of whether gravitation is included in the calculation.
Taking $E$ differentially close to $E_0$, we read off the one-loop
$\beta$ function 
\begin{equation}
\beta(\g,E)=
     -  \frac{b_0}{(4\pi)^2}\g^3-3\frac{\kappa^2}{(4\pi)^2}\g E^2 .
\label{e3.22}
\end{equation}
Using $\kappa^2=16\pi/\Mp^2$, the unknown 
coefficient in Equation (\ref{beta}) is now determined to be 
$ a_0 =-3/\pi\approx -0.95$.

{\it Comments}.---The magnitude $|a_0| \approx 1$ indicates that $\Mp=\sqrt{\hbar c^3/\Gn}$ does indeed give a fair estimate of the energy scale for onset of quantum gravity, with no large numerical factors, for the problem considered here. 

At energy scales a few orders of magnitude below $\Mp$ the discussion that led to Eqs. (\ref{ecorr}) and (\ref{gcorr}) is valid, so the negative sign of $a_0$ slightly increases $E_{\rm U}$ and slightly weakens the value of the unified coupling. This helps to close the gap between the unification scale and the Plank scale.

Gravitational corrections will cause gauge couplings
to run even in theories that in themselves are exactly conformal invariant, that is when $b_0=0$.  
Two notable examples
in four dimensions are  pure $U(1)$ electromagnetism  and $N=4$ 
Super-Yang-Mills \cite{Mandelstam:1982cb,Grisaru:1980nk,Jones:1977zr}. For 
these theories, the exponential integral
in Equation (\ref{formalSolution}) has zero coefficient, so we are left with 
\begin{equation}
e^{-a_0 E^2 / \Mp^2}g^2(E) = {\rm constant}
\label{notQuiteConformal}
\end{equation}
That is, the coupling runs down from its infrared value as a Gaussian with a width of order $\Mp$. For the pure $U(1)$ case, this Gaussian running represents the renormalized coupling strength of photons to non-dynamical or heavy sources, and would be the dominant---but still negligible---source of running in QED far below electron-positron threshold.
For the theoretical ``application'' of Eq. (\ref{notQuiteConformal}) to $N=4$ Super-Yang-Mills, and in the context of unification, it would be logical to include the contribution of gravitino-gluino loops to $a_0$, but we have not calculated that here.  

\begin{figure}
\psfrag{g}{$\g(E)$}
\psfrag{log[10](E/GeV)}{$\log_{10}(E/{\rm GeV})$}
\includegraphics[width=2.75 in, height=2.0in]{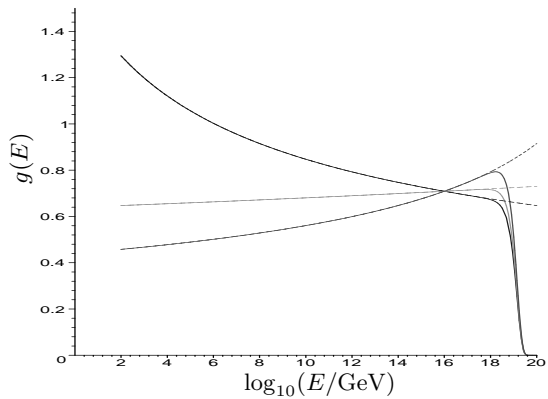}
\caption{\label{fig2} When gravity is ignored, the three gauge couplings of a model theory
evolve as the inverse logarithm of $E$ at one-loop order
(dashed curves).  Initial values at 
100 GeV were set so that the curves exactly intersect at 
approximately $10^{16}$~GeV. When gravity is included at one-loop (solid curves), the couplings remain unified near 
$10^{16}$ GeV, but evolve rapidly towards weaker coupling at high $E$.}
\end{figure}

The negative sign of $a_0$ also signifies that the gravitational correction works in the direction of asymptotic freedom: it causes the couplings to decrease at large energy.   Of course, its effect only becomes quantitatively important when the energy approaches the Planck scale, and soon after that one loses the right to neglect higher-order graviton exchanges.  
Though neglect of additional corrections is not justified beyond $E\ll\Mp$, it is entertaining to consider some consequences of extrapolating Eq. (\ref{formalSolution}) as it stands to these energies, taking into account $a_0 < 0$.    The integral on the right-hand side converges as $E \rightarrow \infty$, and so Eqn. (\ref{notQuiteConformal}) arises as an asymptotic relation.   Thus, the effective coupling vanishes rapidly beyond the Planck scale, rendering the gauge sector approximately free at these energies.
In Fig. \ref{fig2}, we illustrate some aspects of the preceding discussion pictorially for an example theory with three gauge couplings whose low-energy values are chosen such that the $b_0^i$ determined from the matter sector result in a unification at $E_{\rm U}=10^{16}$ GeV. Obviously such a theory mimics the minimally supersymmetric standard model. 


\begin{acknowledgments}
SPR thanks Ted Baltz for useful discussions. 
This work is supported in part by funds provided by the U.S. Department
of Energy (D.O.E.) under cooperative research agreement DE-FC02-94ER40818.
\end{acknowledgments}


\begin{thebibliography}{99}

\bibitem{'tHooft:1974bx}
G.~'t Hooft and M.~Veltman,
Annales Poincare Phys.\ Theor.\ A {\bf 20}, 69 (1974).

\bibitem{Deser:1974cz}
  S.~Deser and P.~van Nieuwenhuizen,
  Phys.\ Rev.\ D {\bf 10}, 401 (1974).

\bibitem{Deser:1974cy}
  S.~Deser and P.~van Nieuwenhuizen,
  Phys.\ Rev.\ D {\bf 10}, 411 (1974).

\bibitem{Deser:1974xq}
  S.~Deser, H.~S.~Tsao, and P.~van Nieuwenhuizen,
  Phys.\ Rev.\ D {\bf 10}, 3337 (1974).


\bibitem{Donoghue:1994dn}
J.~F.~Donoghue,
Phys.\ Rev.\ D {\bf 50}, 3874 (1994).

\bibitem{Georgi:1973}
H.~ Georgi, H.~Quinn, and S.~Weinberg,
Phys.\ Rev.\ Lett. {\bf 33}, 451 (1974).

\bibitem{Dimopoulos:1981yj}
  S.~Dimopoulos, S.~Raby, and F.~Wilczek,
  Phys.\ Rev.\ D {\bf 24}, 1681 (1981).

\bibitem{Kiritsis:1994ta}
  E.~Kiritsis and C.~Kounnas,
  Nucl.\ Phys.\ {\bf B442}, 472 (1995).

\bibitem{Kiritsis:1997hj}
  E.~Kiritsis, {\it Introduction to superstring theory}, Leuven notes in mathematical and theoretical physics, B9 (Leuven University Press, Leuven, Belgium, 1998). 



\bibitem{calculation}
S.~P.~Robinson,  Ph.D. Thesis, M.I.T., 2005 (available at http://web.mit.edu/spatrick/www/PhDThesis/); S.~P.~Robinson and F.~Wilczek (to be published).

\bibitem{'tHooft:1972fi}
  G.~'t Hooft and M.~J.~G.~Veltman,
  Nucl.\ Phys.\ {\bf B44}, 189 (1972).
 
\bibitem{Popov:1967fd}
  V.~N.~Popov and L.~D.~Faddeev, BITP Report No. ITF-67-036, 1967 (unpublished); also Fermilab Report No. FERMILAB-PUB-72-057-T. 

\bibitem{Mandelstam:1982cb}
  S.~Mandelstam,
  Nucl.\ Phys.\ {\bf B213}, 149 (1983).

\bibitem{Grisaru:1980nk}
  M.~T.~Grisaru, M.~Rocek, and W.~Siegel,
  Phys.\ Rev.\ Lett.\  {\bf 45}, 1063 (1980).

\bibitem{Jones:1977zr}
  D.~R.~T.~Jones,
  Phys.\ Lett.\ {\bf 72B}, 199 (1977).

\end{thebibliography}
\end{document}